\documentclass[a4paper,12pt]{article}

\usepackage{amsmath,
amssymb,
bbm,
feynmp,
graphicx,
hyperref,
siunitx,
slashed,
cite,
subfigure}
\usepackage[usenames,dvipsnames,svgnames]{xcolor}

\usepackage{ulem}

\usepackage{geometry}
\newcommand{\be}{\begin{equation}}
\newcommand{\ee}{\end{equation}}
\newcommand{\ba}{\begin{eqnarray}}
\newcommand{\ea}{\end{eqnarray}}

\newcommand{\eq}[1]{\begin{equation}#1\end{equation}}
\newcommand{\Ref}[1]{Ref.~\cite{#1}}
\newcommand{\Eq}[1]{Eq.~(\ref{#1})}

\begin{document}

\begin{flushright}
SI-HEP-2017-02, QFET-2017-02 \\
NIOBE-2017-2, TUM-HEP-1096/17, MIT-CTP-4919 \\
\end{flushright}

\begin{center}
{
\Large
\bf 
Timelike-helicity $B\to \pi\pi$ form factor from\\[4mm]
light-cone sum rules with dipion distribution amplitudes
}
\end{center}

\vspace{0.3cm}
\begin{center}
{\rm Shan Cheng{$^{\, a}$}, Alexander Khodjamirian{$^{\, a}$} and Javier Virto{$^{\, b,\, c}$}} \\[5mm]
{\it\small
{$^{\, a}$} Theoretische Physik 1, Naturwissenschaftlich-Technische Fakult\"at,\\
Universit\"at Siegen, Walter-Flex-Stra\ss{}e 3, D-57068 Siegen, Germany\\[2mm]
{$^{\, b}$} Physics Department T31, Technische Universit\"at M\"unchen,\\ James Frank-Stra{\ss}e 1, D-85748 Garching, Germany\\[2mm]
{$^{\, c}$} Center for Theoretical Physics, Massachusetts Institute of Technology,\\
77 Mass. Ave., Cambridge, MA 02139, USA
}
\end{center}

\vspace{0.8cm}
\begin{abstract}
\vspace{0.2cm}\noindent
We complete the set of QCD light-cone sum rules for $B\to \pi\pi$ transition form factors,   
deriving a new sum rule for the timelike-helicity form factor $F_t$ in terms of dipion  distribution  amplitudes. 
This sum rule, in the leading twist-2 approximation, is directly related to the pion vector form factor.
Employing a relation between $F_t$ and other $B\to \pi\pi$ form factors we obtain also the longitudinal-helicity form factor $F_0$.
In this way, all four (axial-)vector $B\to \pi\pi$ form factors are predicted from light-cone sum rules with dipion distribution amplitudes.
These results are valid for small dipion masses with large momentum.   
\end{abstract}

\vspace{0.5cm}

{\bf 1.} The aim of this note is to present a solution of the problem outlined in \Ref{Hambrock:2015aor},
concerning the calculation of the timelike-helicity $\bar B^0\to\pi^+\pi^0$ form factor $F_t$ from light-cone sum rules.
In \Ref{Hambrock:2015aor}, the method of QCD light-cone sum rules was applied to calculate the $B\to\pi\pi$ transition form
factors, starting from a particular correlation function expanded near the light-cone in dipion distribution amplitudes (DAs). 
The form factor corresponding to the timelike-helicity of the dilepton and denoted as $F_t$ (for a definition and
helicity basis of $B\to \pi\pi$ form factors see e.g.~\Ref{FFKMvD}) could not, however, 
be consistently obtained from that correlation function because of emerging kinematical singularities. 

Here we obtain a new light-cone sum rule for $F_t$ employing a modified
correlation function:
\ba
\Pi_5(q,k_1,k_2)= i\int\,d^4x \, e^{iq\cdot x}\,
\langle \pi^+(k_1)\pi^0(k_2)|
T\{\bar{u}(x)\,i\,m_b\,\gamma_5 \,b(x),\bar{b}(0)\,i\,m_b\,\gamma_5\,d(0)\} |0\rangle\,,
\label{eq:corr}
\ea
where, instead of the axial-vector $b\to u$ transition current,  
the pseudoscalar one is used. 
The immediate advantage of this choice is that the hadronic matrix element of the pseudoscalar current 
is solely determined by the 
timelike-helicity $B\to \pi\pi$ form factor we are interested in:
\be
-iq^\mu\langle  \pi^+(k_1)\pi^0(k_2)|\bar{u}\gamma_\mu \gamma_5 b| \bar{B}^0(p)\rangle
=\langle  \pi^+(k_1)\pi^0(k_2)|\bar{u}\,i\,m_b\,\gamma_5 \, b| \bar{B}^0(p)\rangle =\sqrt{q^2}F_t(q^2,k^2,\zeta)\,.
\label{eq:hmi}
\ee
Here $k^2=(k_1+k_2)^2$ is the invariant mass squared of the dipion system,
and
\eq{
2\zeta-1= \frac{2\,q\cdot \bar k}{\sqrt{\lambda}}=\beta_\pi(k^2) \cos\theta_\pi\ ,
}
where  $\lambda\equiv m_B^4+q^4+k^4-2m_B^2 q^2-2m_B^2 k^2-2q^2 k^2$ is the K\"all\`en function,
$\beta_\pi(k^2)=\sqrt{1-4m_\pi^2/k^2}$, $\bar k = k_1-k_2$,
and $\theta_\pi$ corresponds to the angle between the 3-momenta of
the neutral pion and the $B$-meson in the dipion 
center-of-mass frame.  
Note that we neglect the light $u$ and $d$ quark masses throughout the paper.

The rest of the derivation follows the procedure explained in detail 
in  Ref.~\cite{Hambrock:2015aor}. First we calculate 
the correlation function 
(\ref{eq:corr}) to the leading order (zeroth order in $\alpha_s$) and in 
the lowest twist-2 approximation.
To this end, we contract the $b$-quark fields 
in the free propagator and perform the integration in Eq.(\ref{eq:corr}). 
The result for the 
correlation function (\ref{eq:corr}), which is by itself an invariant amplitude, reads: 
\eq{
\Pi_5(p^2,q^2,k^2,\zeta)=\sqrt{2}m_b^2 \int_{0}^{1} du\ 
\frac{q\cdot k + u\,k^2}{(q+uk)^2-m_b^2} \ \Phi^{I=1}_{\parallel}(u,\zeta,k^2) \,,
\label{eq:ope}}
in terms of the isospin-one dipion DAs 
introduced and defined in \cite{2pionDA,PolyakovNP,Mueller:1998fv,PolyakovWeiss}.
Importantly, the above expression --as opposed to the correlation function 
considered in Ref.~\cite{Hambrock:2015aor}-- is free from kinematical singularities. 
Also important is that Eq.~(\ref{eq:ope}) only depends on 
the chiral-even dipion~DA: 
\eq{
\langle \pi^+(k_1)\pi^0(k_2)|\bar{u}(x)\gamma_\mu[x,0] d(0) |0\rangle=
-\sqrt{2}k_\mu\int\limits_0^1du e^{iu(k\cdot x)}\Phi^{I=1}_{\parallel}(u,\zeta,k^2)\,,
\label{eq:phipar}
}
which is normalized to the pion vector form factor in the timelike region: 
\eq{
\int\limits_0^1 du\ \Phi^{I=1}_{\parallel}(u,\zeta,k^2)=(2\zeta-1)F_\pi(k^2)\,. 
\label{eq:norm}
}
We also use the double expansion of this DA
in partial waves and Gegenbauer polynomials \cite{PolyakovNP}:
\be
\Phi_{\parallel}^{I=1}(u,\zeta,k^2) = 6u\bar{u} 
\sum_{n=0,2,\cdots}^{\infty} \sum_{\ell=1,3,\cdots}^{n+1} B_{n \ell}^{\parallel}(k^2) \, C_{n}^{3/2}(u-\bar{u})
\, \beta_{\pi}(k^2) P_{\ell}^{(0)}(\cos\theta_\pi)\,,
\label{eq:DAexpan}
\ee
where $\bar{u}\equiv 1-u$, $P_{\ell}^{(m)}$ are the associated Legendre polynomials,
and the normalization of the DA in \Eq{eq:norm}
fixes the coefficient $B_{01}^{\parallel}(k^2)=F_{\pi}(k^2)$.

We now insert the $B$-meson ground state in the correlation 
function (\ref{eq:corr}) and, using the definition of $F_t$ in \Eq{eq:hmi}, 
write down the hadronic dispersion relation in the variable 
$p^2=(q+k)^2$, the square of the momentum transferred to the $B$-meson interpolating current:
\be
\Pi_{5}(p^2,q^2,k^2,\zeta)=\frac{f_B m_B^2 \sqrt{q^2}F_t(q^2,k^2,\zeta)
}{m_B^2-p^2}
+ ...\,,
\label{eq:hadr_disp}
\ee
where $f_B$ is the $B$-meson decay constant and the ellipses denote the contributions of 
radially excited and continuum states with $B$-meson quantum numbers. 
The two remaining steps in the derivation of the light-cone sum rule involve: 
(1) employing the quark-hadron duality approximation with a threshold parameter $s_0^B$ 
and  (2) applying the Borel transformation in the variable $p^2$. 
The resulting sum rule reads:
\be
\sqrt{q^2}F_t(q^2,k^2,\zeta)= 
-\frac{m_b^2}{\sqrt{2}f_Bm_B^2}\int\limits_{u_0}^1\frac{du}{u^2}\ 
e^{\frac{m_B^2 -s(u)}{M^2}}
\big(m_b^2-q^2+u^2k^2\big)\ 
\Phi^{I=1}_{\parallel}(u,\zeta,k^2)\,,
\label{eq:lcsr}
\ee
where $s(u)=(m_b^2 - \bar u q^2 + u\bar u k^2)/u $, and $u_0$ is the solution to $s(u_0)=s_0^B$.

Using the above LCSR  for the form factor $F_t$, 
and the sum rule for the form factor 
$F_{\parallel}$ obtained in Ref.~\cite{Hambrock:2015aor}, 
together with the relation between three form factors 
valid to the same twist-2 accuracy, 
we calculate the longitudinal-helicity form factor:
\be
\sqrt{q^2}F_0(q^2,k^2,\zeta)=\frac{1}{m_B^2-q^2-k^2}\Big[\sqrt{\lambda}\sqrt{q^2}F_t(q^2,k^2,\zeta)+
2\sqrt{k^2}\,q^2\,(2\zeta-1)F_{\parallel}(q^2,k^2,\zeta)\Big]\,.
\label{eq:ffrel}
\ee
Thus, all four $B\to \pi\pi$ form factors in the region of small and intermediate
$q^2$ and small $k^2$ are now accessible from LCSRs with dipion DAs. 

Applying the partial wave expansion to the two form factors under consideration, we write
\be
F_{0,t}(q^2, k^2, \zeta) =
\sum_{\ell=0}^{\infty} \sqrt{2 \ell+1}\ F_{0,t}^{(\ell)}(q^2, k^2)\ P_{\ell}^{(0)}(\cos\theta_\pi)\, .
\label{eq:partwav}\\
\ee
Substituting Eqs.~(\ref{eq:DAexpan}) and (\ref{eq:partwav}) into (\ref{eq:lcsr}) and (\ref{eq:ffrel}),
multiplying both sides by $P_{\ell'}^{(0)}(\cos{\theta_{\pi}})$ 
and integrating over $\cos{\theta_{\pi}}$, we obtain the $\ell$-th partial wave contribution 
to the $\bar{B}^0\to \pi^{+}\pi^{0}$ form factor (note that only odd partial 
waves $\ell=1,3,5,...$ 
contribute for the isovector dipion state):
\begin{eqnarray}
\sqrt{q^2}F_{t}^{(\ell)}(q^2, k^2) &=& 
- \frac{6m_b^2}{\sqrt{2} f_B m_B^2} \frac{\beta_{\pi}(k^2)}{\sqrt{2\ell+1}} \nonumber \\
&& \times
\sum_{\binom{n=\ell-1}{n\ even}}^{\infty}  B_{n\ell}^{\parallel}(k^2) \,
\int_{u_0}^{1} \frac{du}{u}\,\bar{u}\, e^{\frac{m_B^2 -s(u)}{M^2}}
(m_b^2-q^2+u^2k^2) \, C_{n}^{3/2}(u-\bar{u})\ ,
\label{eq:lcsrs-Ft}\\[2mm]
\sqrt{q^2}F_{0}^{(\ell)}(q^2, k^2) &=&
\frac{\sqrt{\lambda}\,  \sqrt{q^2}}{m_B^2-q^2-k^2} 
F_{t}^{(\ell)}(q^2, k^2)  
+
\frac{\sqrt{k^2}\,q^2\, \beta_{\pi}(k^2)}{m_B^2-q^2-k^2} 
\sum_{\ell'=1}^{\infty}  \, I_{\ell \ell'} \, F_{\parallel}^{( \ell')}(q^2, k^2),
\label{eq:lcsrs-F0}
\end{eqnarray}
where 
\begin{equation}
I_{\ell\ell'} = \sqrt{2 \ell +1}\sqrt{2 \ell'+1}
\int_{-1}^{1} d\cos{\theta_{\pi}}\ \frac{\cos{\theta_{\pi}}}{\sin{\theta_{\pi}}}
\, P_{\ell}^{(0)}(\cos{\theta_{\pi}}) P_{ \ell'}^{(1)}(\cos{\theta_{\pi}}).
\label{llp}
\end{equation}
Eqs.(\ref{eq:lcsrs-Ft}) and (\ref{eq:lcsrs-F0})
complement the ones for the partial waves of the 
form factors $F_{\parallel,\perp}^{(\ell)}$ obtained in Ref~\cite{Hambrock:2015aor}.

\bigskip


{\bf 2.}  To assess the dominant $\rho$-resonance contribution to the $P$-wave of $B\to \pi\pi$ timelike-helicity form factor,
we follow Refs.~\cite{Hambrock:2015aor,Cheng:2017smj} and relate the  form factor $F_t$ to the corresponding $B\to \rho$ form factor  $A_0(q^2)$ 
(defining the $B\to \rho$ form factors as in Refs.\cite{BBFFBrho,Ball:2004rg}) by means of a resonance model:
\be
\sqrt{q^2} F_{t}^{(\ell=1)}(q^2, k^2) =
-\frac{\sqrt{\lambda} \ \beta_{\pi}(k^2)}{\sqrt{3}}
\frac{g_{\rho\pi\pi}m_{\rho}A_{0}(q^2)}{[m_{\rho}^2-k^2-i\sqrt{k^2}\,\Gamma_{\rho}(k^2)]} + \cdots \,, 
\label{eq:Brho}\\
\ee
where the ellipsis  denotes the contributions of excited resonances,
and the energy-dependent width $\Gamma_\rho(k^2)$ (see definition in Eq.~(36) of 
Ref~\cite{Hambrock:2015aor}) effectively takes into account the 
two-pion  mixing  with the $\rho$.  
For $A_0(q^2)$ we derive a LCSR in terms of the $\rho$-meson DA  (in the zero-width approximation) taking for consistency the twist-2 LO approximation. 
This sum rule is similar to the sum rules derived in Refs.~\cite{BBFFBrho,Ball:2004rg}, where one can also find the necessary details on the $\rho$-meson~DAs. We find: 
\be
A_{0}(q^2) = \frac{m_b^2f_\rho}{2 f_B m_B^2} \, 
\int_{u_0}^{1} \frac{du}{u} \, 
\exp\Big(\frac{m_B^2}{M^2}-\frac{m_b^2-\bar{u}\, q^2+ u\bar{u}\, m_\rho^2}{u\,M^2}\Big)\,
 \phi^{\rho}_\parallel(u) \, ,
\label{eq:A0lcsr}
\ee
containing the chiral-even $\rho$-meson~DA~$\phi^\rho_\parallel(u)$, normalized
to the $\rho$-meson decay constant $f_\rho$.  

Note that  if we substitute the LCSRs (\ref{eq:lcsrs-Ft}) for $F_{t}^{(\ell=1)}(q^2, k^2)$ and (\ref{eq:A0lcsr}) for $A_{0}(q^2)$ to the left-hand and
right-hand sides of the one-resonance approximation (\ref{eq:Brho}), respectively, 
and use for simplicity the asymptotic 2-pion and $\rho$-meson DAs, that is:
$B_{n>0,\ell}=0$ and $\phi^\rho_\parallel(u)=6u(1-u)$, the resulting relation 
will restore  $B_{01}(k^2)=F_\pi(k^2)$ in the form of the $\rho$-meson 
contribution to the pion form factor {({\it e.g.,} Eq.(25) in \Ref{Cheng:2017smj})}.
However, this is valid only up to power 
corrections of ${\cal O}(q^2/m_b^2)$, ${\cal O}(k^2/m_b^2)$ and ${\cal O}(\Delta/m_b)$,
where we rescale the effective threshold as
$s_0^B=(m_b+\Delta)^2$. The 
latter correction originates from 
a global factor of $1/u$ in the integrand.

\bigskip


\begin{figure}
\begin{center}
\includegraphics[width=0.48\textwidth]{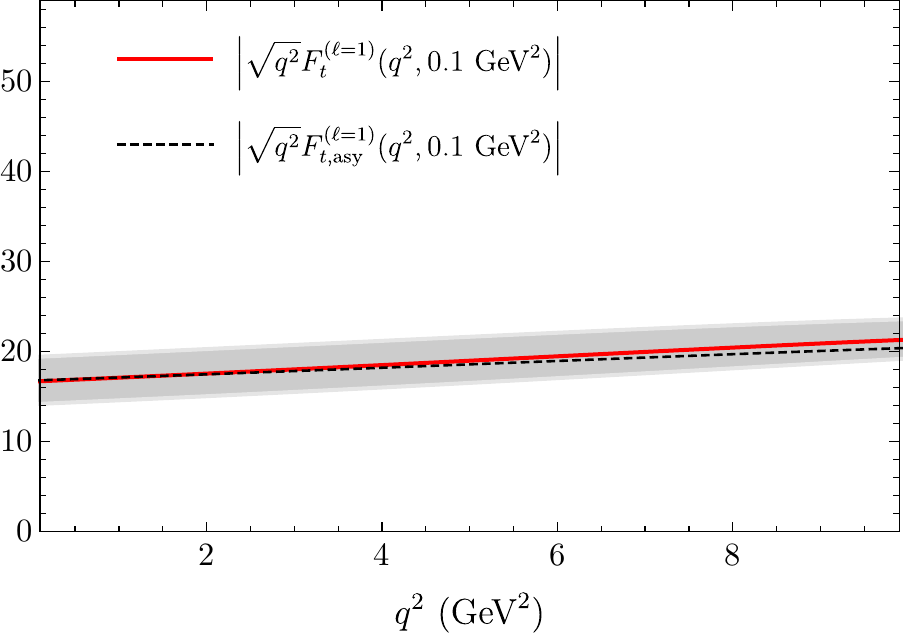}
\hspace{3mm}
\includegraphics[width=0.48\textwidth]{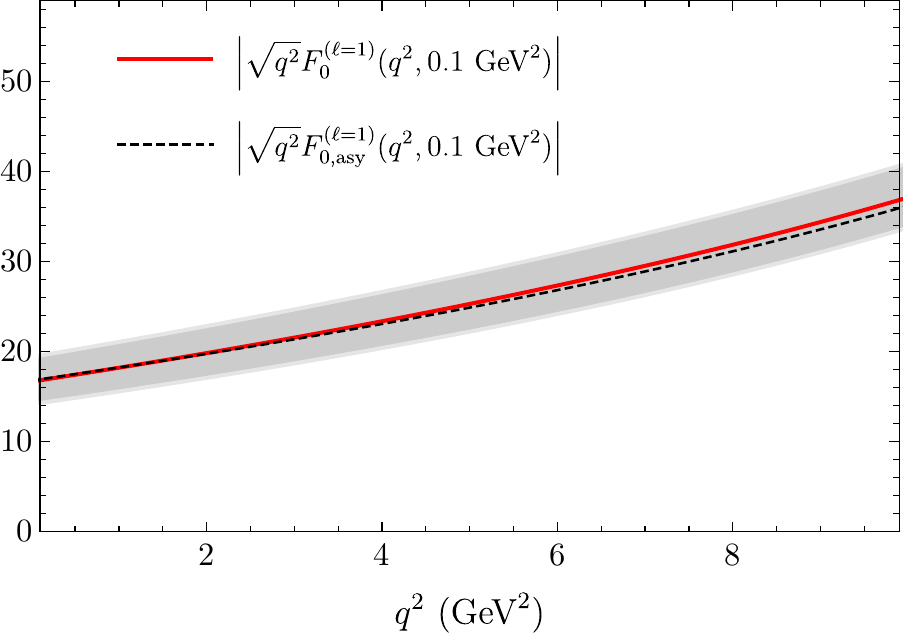}
\end{center}
\vspace{-0.5cm}
\caption{LCSR predictions for the absolute values of the timelike- and longitudinal-helicity
\mbox{$P$-wave} $B\to \pi\pi$ form factors $\sqrt{q^2}F^{(\ell=1)}_{t,0}(q^2, k^2=0.1~\textrm{GeV}^2)$. 
Solid (dashed) curves are the central values with the nonasymptotic 
(asymptotic) dipion DAs. Shaded bands show the estimated theoretical uncertainties.}
\label{fig:1}
\end{figure}

{\bf 3.}  For the numerical analysis  of the new sum rule (\ref{eq:lcsr})
we adopt the same input as in Ref.~\cite{Hambrock:2015aor} in particular:
the $b$-quark mass, decay constant of $B$, Borel parameter
range and the duality threshold. The nonperturbative 
universal input encoded in the functions $B^{\parallel}_{n\ell}(k^2)$
entering the Gegenbauer expansion of the DA
are taken from the instanton model used in \Ref{PolyakovNP}  and 
are listed in Eqs.~(6.7)-(6.12) there. 
These estimates are only valid near the dipion threshold $k^2\sim 4m_\pi^2$
\footnote{For numerical illustration we will take $k^2=0.1 \, \textrm{GeV}^2$,
slightly above the two-pion threshold,  
since the phase-space factor $\beta_{\pi}(k^2)$ in Eq.~(\ref{eq:lcsrs-Ft}) makes the
form factor vanish at threshold.}. 
Hence, we are able to predict the $q^2$-dependence of the 
form factors $F^{(\ell)}_{t,0}(q^2,k^2\sim 4m_\pi^2)$
in the region of large recoil $0\leq q^2\leq 10-12$ GeV$^2$, 
where one can trust the light-cone expansion of the correlation function. 
The results for the $P$-wave form factors are plotted in Fig.~\ref{fig:1}, 
where only the uncertainties from the sum rule parameters 
$s^B_0$ and $M^2$ are shown.

\begin{figure}
\begin{center}
\includegraphics[width=0.65\textwidth]{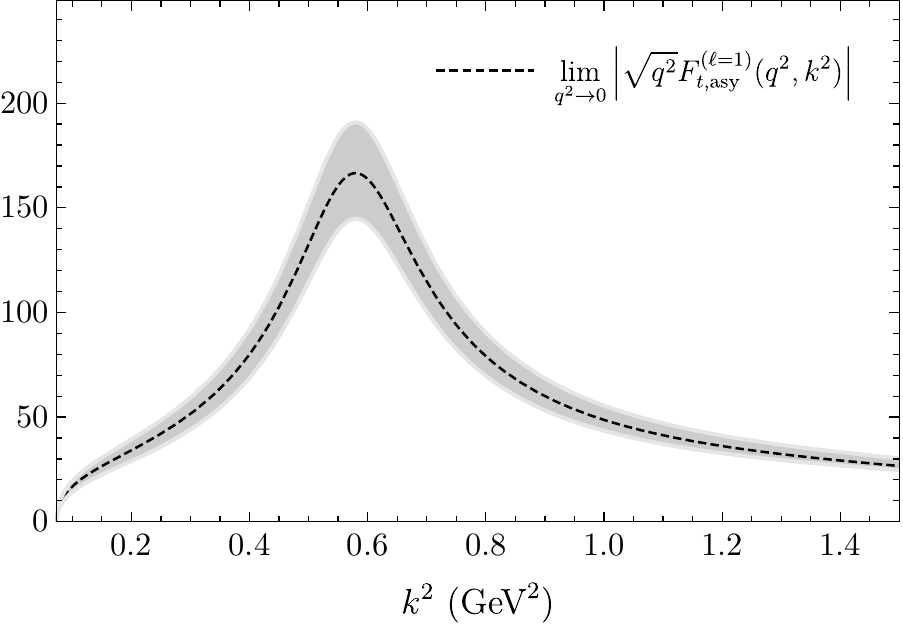}
\end{center}
\vspace{-0.5cm}
\caption{Absolute value of the timelike-helicity \mbox{$P$-wave}
$B\to \pi\pi$ form factor $\sqrt{q^2}F^{(\ell=1)}_{t}(q^2,k^2)$ at $q^2=0$ obtained
from LCSR with the asymptotic dipion DA, using the data \cite{Fujikawa:2008ma}
on the pion form factor.}
\label{fig:2}
\end{figure}

We find that the higher partial waves are also strongly suppressed
in $F_t$ as in the other form factors considered in 
Ref.~\cite{Hambrock:2015aor}. E.g., the 
ratio of $\ell=3$ to $\ell=1$ contributions to $F_t(q^2,4 m_\pi^2)$ 
is smaller than $5\%$ at all accessible $q^2$.
The contribution of nonasymptotic terms are also small, 
as can bee seen by setting to zero all $B^{\parallel}_{n\ell}$ except 
$B^{\parallel}_{01}$ 
(see the dashed curves in Fig.~\ref{fig:1}). 
This allows us to predict the $P$-wave form factor
also at larger $k^2$, including  the $\rho$-resonance  region and even beyond,
provided we use for $B^{\parallel}_{01}(k^2)$ the accurate data for the pion form factor $F_\pi(k^2)$ provided by the Belle collaboration~\cite{Fujikawa:2008ma}. 
Our result for $\lim_{q^2\to 0}\sqrt{q^2}F^{(\ell=1)}_t(q^2,k^2)$, as a function of 
dipion invariant mass squared is shown in Fig.~\ref{fig:2}.

We have also calculated the $B\to \rho$ contribution to the form factor 
$F^{(\ell=1)}_t(q^2,k^2)$ using \Eq{eq:Brho} and find  that the remaining 
resonant and continuum contributions to this form factor can amount 
up to $20\%$ in the small $k^2$ region, 
in agreement with the findings of Refs.~\cite{Hambrock:2015aor,Cheng:2017smj} for the
other $B\to \pi\pi$ form factors.

\bigskip

{\bf 4.} Concluding, we have completed the set of LCSRs with dipion DAs for the  $B\to \pi\pi$ form factors. These sum rules are complementary to the ones derived in \Ref{Cheng:2017smj} in terms of $B$-meson DAs, and complementary to the derivations from dispersion theory~\cite{Kang:2013jaa}, and to the calculations at large $k^2$~\cite{Boer:2016iez}.
The form factor $F_t$ obtained here, while not contributing to the semileptonic $B\to \pi\pi \ell \nu$ rate in the massless lepton approximation,
plays an important role in the factorization formula for nonleptonic $B\to \pi\pi\pi$ decays~\cite{Krankl:2015fha,Virto:2016fbw,Klein:2017xti}. Further improvements of the sum rules presented here and in Refs.~\cite{Hambrock:2015aor,Cheng:2017smj} require the
inclusion of higher-twists and NLO corrections, but most importantly -- for the ones derived here and in \Ref{Hambrock:2015aor} --
a better knowledge of dipion DAs and their Gegenbauer coefficients $B_{n\ell}$.

\section*{Acknowledgments}
This work is supported by the DFG Research Unit FOR 1873
``Quark Flavour Physics and Effective Theories'', 
contract No KH 205/2-2. 
J.V. acknowledges funding from the Swiss National Science Foundation,
from Explora project FPA2014-61478-EXP,
and from the European Union's€š Horizon 2020 research and innovation programme under the Marie Sklodowska-Curie grant agreement No 700525 `NIOBE'.


\begin{thebibliography}{100}

\bibitem{Hambrock:2015aor} 
  C.~Hambrock and A.~Khodjamirian,
 ``Form factors in $\bar B^0 \to \pi\pi\ell\bar\nu_\ell$ from QCD light-cone sum rules,''
  Nucl.\ Phys.\ B {\bf 905}, 373 (2016),
  arXiv:1511.02509 [hep-ph].


\bibitem{FFKMvD}
  S.~Faller, T.~Feldmann, A.~Khodjamirian, T.~Mannel and D.~van Dyk,
  ``Disentangling the Decay Observables in $B^- \to \pi^+\pi^-\ell^-\bar\nu_\ell$,''
  Phys.\ Rev.\ D {\bf 89} (2014) 014015,
  arXiv:1310.6660 [hep-ph].

\bibitem{2pionDA}
  M.~Diehl, T.~Gousset, B.~Pire and O.~Teryaev,
  ``Probing partonic structure in $\gamma^* \gamma \to \pi \pi$ near threshold,''
  Phys.\ Rev.\ Lett.\  {\bf 81} (1998) 1782,
  hep-ph/9805380.


\bibitem{PolyakovNP}
  M.~V.~Polyakov,
  ``Hard exclusive electroproduction of two pions and their resonances,''
  Nucl.\ Phys.\ B {\bf 555} (1999) 231,
  hep-ph/9809483.

\bibitem{Mueller:1998fv}
  D.~M\"uller, D.~Robaschik, B.~Geyer, F.~M.~Dittes and J.~Ho\u{r}ej\u{s}i,
 ``Wave functions, evolution equations and evolution kernels from light ray operators of QCD,''
  Fortsch.~Phys.~{\bf 42}~(1994) 101,
  hep-ph/9812448.


\bibitem{PolyakovWeiss}
  M.~V.~Polyakov and C.~Weiss,
 ``Two pion light cone distribution amplitudes from the instanton vacuum,''
  Phys.\ Rev.\ D {\bf 59} (1999) 091502,
  hep-ph/9806390.


\bibitem{Cheng:2017smj} 
  S.~Cheng, A.~Khodjamirian and J.~Virto,
  ``$B\to\pi\pi$ Form Factors from Light-Cone Sum Rules with $B$-meson Distribution  Amplitudes,''
  JHEP {\bf 1705}, 157 (2017),
  arXiv:1701.01633 [hep-ph].


\bibitem{BBFFBrho}
  P.~Ball and V.~M.~Braun,
 ``Use and misuse of QCD sum rules in heavy to light transitions: The Decay $B \to \rho e \nu$ reexamined,''
  Phys.\ Rev.\ D {\bf 55} (1997) 5561,
  hep-ph/9701238.

\bibitem{Ball:2004rg} 
  P.~Ball and R.~Zwicky,
  ``$B_{d,s} \to  \rho, \omega, K^*, \phi$ decay form-factors from light-cone sum rules revisited,''
  Phys.\ Rev.\ D {\bf 71}, 014029 (2005),
  hep-ph/0412079.

\bibitem{Kang:2013jaa} 
  X.~W.~Kang, B.~Kubis, C.~Hanhart and U.~G.~Mei§ner,
  ``$B_{l4}$ decays and the extraction of $|V_{ub}|$,''
  Phys.\ Rev.\ D {\bf 89}, 053015 (2014),
  arXiv:1312.1193 [hep-ph].

\bibitem{Boer:2016iez} 
  P.~B\"oer, T.~Feldmann and D.~van Dyk,
  ``QCD Factorization Theorem for $B \to \pi\pi\ell\nu$ Decays at Large Dipion Masses,''
  JHEP {\bf 1702}, 133 (2017),
  arXiv:1608.07127 [hep-ph].

\bibitem{Krankl:2015fha} 
  S.~Kr\"ankl, T.~Mannel and J.~Virto,
  ``Three-Body Non-Leptonic B Decays and QCD Factorization,''
  Nucl.\ Phys.\ B {\bf 899}, 247 (2015),
  arXiv:1505.04111 [hep-ph].
  
\bibitem{Virto:2016fbw} 
  J.~Virto,
  ``Charmless Non-Leptonic Multi-Body B decays,''
  PoS FPCP {\bf 2016}, 007 (2017),
  arXiv:1609.07430 [hep-ph].

\bibitem{Klein:2017xti} 
  R.~Klein, T.~Mannel, J.~Virto and K.~K.~Vos,
  ``CP Violation in Multibody $B$ Decays from QCD Factorization,''
  arXiv:1708.02047 [hep-ph].


\bibitem{Fujikawa:2008ma}
  M.~Fujikawa {\it et al.} [Belle Collaboration],
 ``High-Statistics Study of the $\tau^- \to \pi^- \pi^0 \nu_\tau$ Decay,''
  Phys.\ Rev.\ D {\bf 78} (2008) 072006,
  arXiv:0805.3773 [hep-ex].




\end{thebibliography}
\end{document}